\documentclass{jpsj-suppl}
\usepackage{txfonts} 

\usepackage{wrapfig}
\newcommand{\bra}[1]{\langle {#1} |}
\newcommand{\ket}[1]{| {#1} \rangle}

\title{Quenching of $N=28$ shell gap and \\
a low-lying quadrupole mode \\
in the vicinity of neutron-rich $N=28$ isotones}

\author{Shuichiro \textsc{Ebata}$^{1}$ and Masaaki \textsc{Kimura}$^{2}$}

\inst{$^{1}$Meme Media Laboratory, Hokkaido University, Sapporo 060-0813, Japan\\
$^{2}$Department of Physics, Hokkaido University, Sapporo 060-0810, Japan}

\email{ebata@nucl.sci.hokudai.ac.jp}

\recdate{July 15, 2013}

\abst{
We investigate the low-lying 2$^+$ states of $N=28$ isotones 
($^{48}$Ca, $^{46}$Ar, $^{44}$S and $^{42}$Si) with using 
the canonical-basis time-dependent Hartree-Fock-Bogoliubov theory 
in which the pairing is taken into account self-consistently. 
The quadrupole mode with very small excitation energies 
emerges in $^{46}$Ar, $^{44}$S and $^{42}$Si 
due to the strong quadrupole correlations triggered 
by the quenching of the $N=28$ shell gap. 
The importance of the quadrupole correlations between the protons and neutrons, and the role of
the pairing correlation to reinforce the lowest 2$^+$ mode are discussed. 
It is also shown that the $B$(E2$\uparrow$) values of $^{48}$Ca and $^{46}$Ar 
are plausibly explained by our calculations.
}

\kword{Shell quenching, 2$^+$ state, Pairing, Time-dependent mean field theory}

\begin{document}
\maketitle
\subsection*{\rm \bf Introduction}
The low-energy excited modes are quite sensitive to the underlying shell structure 
and the pairing correlations, and hence, the quenched magic shell gaps 
of unstable nuclei should generate unique collective modes.
One of the interesting example is the $N$=28 shell gap that is known to be quenched
in the vicinity of $^{44}$S, and has been paid considerable experimental \cite{hS96}
and theoretical attention \cite{RCF97}.

Since there is a $\Delta l$=2 difference of orbital angular momentum between 
$f_{7/2}$ and $p_{3/2}$ states, the quench of $N$=28 shell gap will 
induce the strong quadrupole correlation in the low-lying state. 
Furthermore, the protons in Si, S and Ar isotopes
occupy the middle of the $sd$-shell, and hence, 
the strong quadrupole correlation should also exist in the proton side.
Therefore, when the $N$=28 shell gap is quenched, 
the strong quadrupole correlations among protons and neutrons will be ignited and 
can be expected to lead to a variety of the excitation modes.
Indeed, various exotic phenomena such as the shape transition in Si and S isotopes 
and the shape coexistence is theoretically suggested \cite{mK13}. 

We investigate the low-lying quadrupole excitation (E2) modes in $^{46}$Ar, $^{44}$S and $^{42}$Si 
generated by the strong quadrupole correlation between protons and neutrons. 
To access the E2 modes, we apply 
the canonical-basis time-dependent Hartree-Fock-Bogoliubov (Cb-TDHFB) theory \cite{EN10} 
which can be successfully applied to the study of 
the dipole and quadrupole modes of even-even isotopes \cite{EN10, SL13, EN14}. 

\subsection*{\rm \bf Formulation}
\subsection*{Cb-TDHFB equations}
The Cb-TDHFB can describe self-consistently the dynamical effects of pairing correlation 
which has a significant role to generate the low-lying quadrupole strength 
as reported in Ref. \cite{YG04}.  
By assuming the diagonal form of pairing functional, 
the Cb-TDHFB equations are derived from the full TDHFB equation 
represented in the canonical basis $\{ \phi_l(t), \phi_{\bar l}(t)\}$ 
which diagonalize a density matrix.  
The Cb-TDHFB equations describe the time-evolution of the canonical pair 
$\{ \phi_l(t),\phi_{\bar l}(t) \}$, its occupation probability $\rho_l(t)$ 
and pair probability $\kappa_l(t)$,  
\begin{eqnarray}
i \hbar \frac{\partial}{\partial t}\ket{\phi_l(t)}& &\hspace{-7mm} = [\hat{h}(t) - \eta_{l}(t)]\ \ket{\phi_l(t)}, \ \ \ \ 
i \hbar \frac{\partial }{\partial t}\ket{\phi_{\bar l}(t)}  = [\hat{h}(t) - \eta_{\bar{l}}(t)]\ \ket{\phi_{\bar l}(t)},  \nonumber \\
i \hbar \frac{\partial \rho_{l}(t)}{\partial t}& &\hspace{-7mm} = \kappa_{l}(t)\Delta_{l}^{\ast}(t) - \kappa_{l}^{\ast}(t)\Delta_{l}(t),  \nonumber \\
i \hbar \frac{\partial \kappa_{l}(t)}{\partial t}& &\hspace{-7mm}= [\eta_{l}(t) + \eta_{\bar l}(t) ]\ \kappa_{l}(t) + \Delta_{l}(t) [2\rho_{l}(t) - 1], 
\label{eq:Cb-TDHFB}
\end{eqnarray} 
where $\eta_l(t)\equiv\bra{\phi_l(t)}h(t)\ket{\phi_l(t)}$, and the $h(t)$ and $\Delta_l(t)$
are the single-particle Hamiltonian and the gap energy, respectively. 
We apply the Skyrme interaction with SkM$^*$ parameter set to $ph$-channel 
and the simple pairing form
\begin{eqnarray}
\Delta_l(t) = \sum_{k>0} G_{kl}\ \kappa_k(t)\ 
\equiv \sum_{k>0} G_0\ f(\varepsilon_k^0)f(\varepsilon_l^0)\ \kappa_k(t),
\label{eq:delta}
\end{eqnarray}
to $pp(hh)$-channel, where $f(\varepsilon)$ is a cutoff function. 
$\varepsilon_l^0, \varepsilon_{\bar l}^0$ mean 
the single-particle energies of the canonical states $\phi_l(t=0), \phi_{\bar l}(t=0)$. 
We choose $G_0, f(\varepsilon^0)$ as a constant value in the real-time evolution.  
The detail form of $f(\varepsilon)$ and evaluation of $G_0$ is used 
as same as Ref.\cite{EN10,TTO96}. 
In accordance with Ref. \cite{NY05}, we also apply the absorbing boundary condition 
to eliminate unphysical modes. 
The canonical basis $\phi_l(\vec{r},\sigma; t) = \langle \vec{r},\sigma | \phi_l(t) \rangle$ with $\sigma=\pm 1/2$ is
represented in the three-dimensional coordinate discretized in a square mesh of 1 fm 
in a sphere with radius of 18 fm including 6 fm for the absorbing potential whose 
the depth is $-3.75$ MeV.  

\subsection*{Linear response calculation with a time-dependent scheme}
We use the linear response calculation to investigate excited modes. 
A numerical small amplitude is added to the ground state, 
and the information of excited states is extracted from the density fluctuations. 
To induce E2 modes, we add a weak instantaneous external field 
$V_{\rm ext}(\vec{r},t)=\eta\hat{F}_K(\vec{r}) \delta(t)$ to initial states 
of the time evolution. 
Here the quadrupole external field acting on proton, neutron, isoscalar (IS) and
isovector (IV) channels are given as 
$\hat{F}_{K} \equiv  (\frac{1\mp\tau_z}{2},1{\rm\ or\ }\tau_z)\otimes (r^2 Y_{2K} + r^2
Y_{2-K})/\sqrt{2(1+\delta_{K0})}$.  
The amplitude of the external field is so chosen to be a small number
$\eta=1 \sim 3\times10^{-3}$ fm$^{-2}$ to guarantee the linearity.
The strength function $S(E;\hat{F}_K)$ in each channel is obtained 
through the Fourier transformation of the time dependent expectation value 
${\cal F}_K(t) \equiv \langle \Psi(t)| \hat{F}_K | \Psi(t) \rangle$:  
\begin{eqnarray}
S(E;\hat{F}_K) \equiv \sum_{n} |\langle \Psi_n | \hat{F}_K | \Psi_0 \rangle |^2  \delta(E_n - E) 
=  \frac{-1}{\pi \eta} {\rm Im} \int_{0}^{\infty} \!\! \{ {\cal F}_K(t) - {\cal F}_K(0) \} e^{i(E+i\Gamma/2)t} dt,
 \nonumber
\end{eqnarray}
where $| \Psi_0 \rangle,\ | \Psi_n \rangle$  and $| \Psi (t) \rangle$ are the ground and
excited states and a time-dependent many body wave function represented in the canonical
form, respectively. $\Gamma$ is a smoothing parameter set to 1 MeV. 

We also performed unperturbed calculations in which $h(t)$ in Eq.(\ref{eq:Cb-TDHFB}) 
is replaced with the static single particle Hamiltonian $h(t=0)$ 
computed with the ground state density.
By comparing the results obtained by the fully self-consistent 
and unperturbed calculations, we investigate the effects of the residual interaction 
and the collectivity of the excited states. 
\subsection*{\rm \bf Results} 
\begin{figure}[t]
\begin{center}
\includegraphics[keepaspectratio, width=5.5cm, clip, angle=-90]{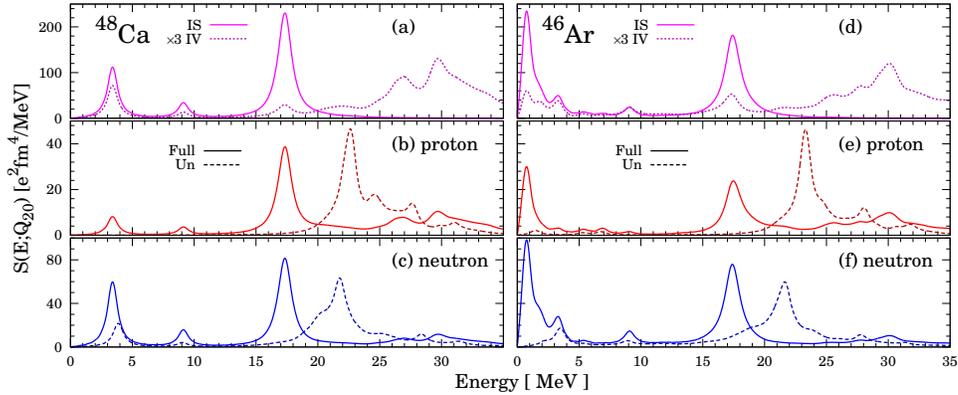}
\caption{
(Color on-line) Strength functions of quadrupole modes of $^{48}$Ca and $^{46}$Ar. 
  The strength functions in the IS (solid line) and
  IV (dotted line) channels are shown in the panels (a) and (d), while those in the
  proton and neutron channels are shown in (b), (e) and (c), (f). The solid and
  dashed lines in the  panels (b), (c), (e) and (f) compare the fully self-consistent and 
  unperturbed results.}
\label{fig:48Ca46ArQ}
\end{center}
\end{figure}
Figure 1 shows strength functions of quadrupole vibrational modes (IS, IV, proton, neutron) 
for $^{48}$Ca and $^{46}$Ar.
The strength functions in the IS and IV channels shown in Fig. \ref{fig:48Ca46ArQ} 
(a) have two peaks at 3.40 and 9.12 MeV in addition to the IS giant quadrupole resonance (GQR) 
at 17 MeV and the IVGQR having broad distribution around 30 MeV. 
The properties of these two peaks below 10 MeV become clear by comparing 
the results in the proton and neutron channels  (Fig. \ref{fig:48Ca46ArQ} (b) and (c)). 
Their strengths in the neutron channel are much larger than those in the proton
channel  showing the dominance of neutron excitation and they are understood as 
the neutron single-particle excitations across the $N$=28 shell gap.
Indeed, they correspond to the 3.86 and 9.06 MeV peaks in the unperturbed results (dashed line) 
that are the neutron single-particle excitations of $f_{7/2}$ ($\varepsilon_{f_{7/2}}$= $-$10.41 MeV) 
$\to$ $p_{3/2}$ ($\varepsilon_{p_{3/2}}$= $-$6.55 MeV) and $f_{5/2}$ ($\varepsilon_{f_{5/2}}$= $-$1.36 MeV), 
respectively. 
On the other hand, in the proton channel, there is no peak below 10 MeV in the unperturbed result, 
since the proton excitation with $\Delta l =2$ costs much larger energy due to the $Z=20$ shell closure. 
In proton channel, the peaks of the fully self-consistent results seem to be induced due to the neutron $2^+$ excitation. 

$^{46}$Ar has different nature of the ground and low-lying $2^+$ states. 
The ground state of $^{46}$Ar is also spherical but superfluid phases appear 
in both of proton ($\Delta^{\rm p}=1.16$ MeV) and neutron ($\Delta^{\rm n}=1.77$ MeV). 
And if the proton pairing is switched off the ground state is oblately deformed ($\beta=-0.15$). 
These results imply the weaker neutron-magicity in $^{46}$Ar than in $^{48}$Ca. 
Actually, the calculated $N$=28 shell gap is slightly reduced in $^{46}$Ar (3.5 MeV) 
compared to that in $^{48}$Ca (3.86 MeV).  
The unperturbed strengths in the proton and neutron channels 
in the Fig. \ref{fig:48Ca46ArQ} (e) and (f) are quite similar to those of $^{48}$Ca 
except for minor differences; 
(1) the reduction of the peak energies below 10 MeV in the neutron channel 
due to the quenching of the $N$=28 shell gap, 
and (2) the very weak strength distributed below 10 MeV in the proton channel 
that are generated by the proton hole-states in $sd$-shell 
and fluctuated by the pairing correlation. 
In the fully self-consistent results, very strong peaks emerge around 1 MeV 
in all channels as shown in Fig. \ref{fig:48Ca46ArQ} (d)-(f). 

The same mechanism also applies to other $N$=28 isotones $^{44}$S and $^{42}$Si.
In the present calculation, $^{44}$S has slightly deformed shape ($\beta=0.08$), while 
$^{42}$Si has oblate shape ($\beta=-0.19$).
Deformation of these nuclei splits the strength functions in the $K=0, 2$ modes, 
and makes their strength distributions more complicated than those in $^{46}$Ar. 
However, we could still identify very low-lying peaks which correspond to 
a couple of $2^+$ states with enhanced collectivity.

To analyze the peak structure, we fit the strength function $S(E;\hat{F}_K)$
below 25 MeV by the sum of Lorentzian $f_k(E;\hat{F}_{K})$:  
\begin{eqnarray} 
\label{eq:fit}
S(E;\hat{F}_K) 
\simeq \sum_{k} f_k(E;\hat{F}_{K})  
\equiv \sum_{k} \frac{a_k^{\tau} (\Gamma/2)^2}{[E-E(2_k^+)]^2 + (\Gamma/2)^2},
\end{eqnarray}
where $a_k^\tau$ is the amplitude of fitted Lorentzian and $\Gamma = 1$ MeV corresponding to smoothing
parameter in Eq. (\ref{eq:Cb-TDHFB}). 
The reduced transition probabilities in the proton and neutron channels, 
that are denoted as $B_k$(E2$\uparrow$) in the following, 
are evaluated by integrating $f_k(E;\hat{F}_{K})$ for each $2_k^+$ state, 
\begin{eqnarray} 
B_k({\rm E2}\!\uparrow ) \equiv 
\! \sum_{K} |\langle 2_k^+| \hat{Q}_{2K}^{\rm p} |0^+_1 \rangle|^2 
\simeq 5\! \int_0^{\infty} \! f_k(E;\hat{F}_{K=0})\ dE.
\end{eqnarray}
This evaluation is reasonable for spherical nuclei such as $^{48}$Ca and $^{46}$Ar. 

\begin{wrapfigure}[15]{r}{70mm}
\begin{center}
\ \\[-12mm]
\includegraphics[keepaspectratio, width=5cm, angle=-90]{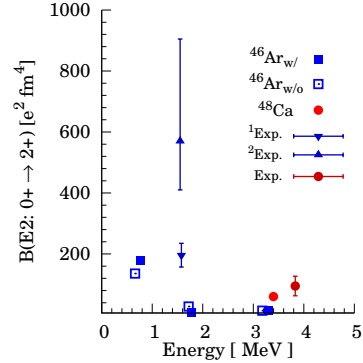}
\caption{(Color on-line) 
Evaluated and experimental $B$(E2$\uparrow$) values for $^{48}$Ca and $^{46}$Ar. 
The experimental data are taken from Ref.\cite{hS96, RNT01, Me10}.}
\label{fig:BE2}
\end{center}
\end{wrapfigure}
The results of fitting, evaluated $B$(E2$\uparrow$)  below 5 MeV 
are shown in Fig. \ref{fig:BE2}. 
The square and filled circle symbols indicate 
the calculated $B$(E2$\uparrow$) of $^{46}$Ar and $^{48}$Ca, respectively. 
The open square symbols show the results without the neutron pairing. 
By comparing the results with and without the pairing in the neutron channel, 
it turns out that the pairing correlation 
enhances $B$(E2$\uparrow$) \cite{YG04} by about 20\% in this work. 
The symbols with error bar are experimental data, 
but the triangle and under triangle show new and old one.
Experimentally, the 2$_1^+$ state of $^{48}$Ca locates at 3.83 MeV and 
$B$(E2$\uparrow$)=95$\pm32$ $e^2$fm$^4$ \cite{RNT01} 
that are in good agreement with the calculation. 
For $^{46}$Ar, 
two experimental $B$(E2$\uparrow$) values are reported 
in the Coulomb-excitation study $B$(E2$\uparrow$)=196$\pm39$ $e^2$ fm$^4$ at $E(2_1^+)=1.58$ MeV 
and in the lifetime measurement $B$(E2$\uparrow$)=570$_{-160}^{+335}\ e^2$ fm$^4$ \cite{Me10}. 
Our evaluated $B$(E2$\uparrow$) value nicely agrees with that of the Coulomb-excitation study, 
although our energy of the first 2$^+$ state is underestimated, 
but for the value of the lifetime measurement, ours does not at all. 

\subsection*{\rm \bf Summary}
We investigated the low-lying quadrupole vibrational modes of $N$=28
isotones by using Cb-TDHFB theory. 
We pointed out
the importance of quadrupole correlation between protons and neutrons that couples proton
hole states with neutron excitations across $N$=28 shell gap.
It is found that the
quenching of $N$=28 shell gap and the proton holes in the $sd$-shell trigger strong
quadrupole correlation and generate the low-lying $2^+$ states in $^{46}$Ar.

\subsection*{\rm \bf Acknowledgment} 
We would like to thanks Prof. T. Nakatsukasa for giving us the useful comments and advice. 
The calculations have been supported by the high performance computing system 
at Research Center for Nuclear Physics, Osaka University.

\end{document}